\documentclass[pra,twocolumn,superscriptaddress,10pt]{revtex4-1}
\usepackage{graphicx}
\usepackage{dcolumn}
\usepackage{color}
\usepackage{times}
\usepackage{bm}
\usepackage{amssymb}
\usepackage{amsmath}
\usepackage{epsfig}
\usepackage{epstopdf}
\usepackage{dsfont}
\usepackage{subfigure}
\begin{document}
\renewcommand{\thefootnote}{\fnsymbol {footnote}}
	
	\title{\textbf{Inequality relation between entanglement and Bell nonlocality for arbitrary two-qubit states}}
	
	\author{Xiao-Gang Fan}
	\affiliation{School of Physics \& Material Science, Anhui University, Hefei 230601, China}
	
    \author{Zhi-Yong Ding}
	\affiliation{School of Physics \& Material Science, Anhui University, Hefei 230601, China}
	\affiliation{School of Physics and Electronic Engineering, Fuyang Normal University, Fuyang 236037, China}

    \author{Fei Ming}
	\affiliation{School of Physics \& Material Science, Anhui University, Hefei 230601, China}

    \author{Huan Yang}
	\affiliation{School of Physics \& Material Science, Anhui University, Hefei 230601, China}
	\affiliation{Department of Experiment and Practical Training Management, West Anhui University, Lu'an 237012, China}

	\author{Dong Wang}
	\affiliation{School of Physics \& Material Science, Anhui University, Hefei 230601, China}
	
	\author{Liu Ye}
	\email{yeliu@ahu.edu.cn}
	\affiliation{School of Physics \& Material Science, Anhui University, Hefei 230601, China}
	
\begin{abstract}
Entanglement and Bell nonlocality are used to describe quantum inseparabilities. Bell-nonlocal states form a strict subset of entangled states. A natural question arises concerning how much territory Bell nonlocality occupies entanglement for a general two-qubit entangled state. In this work, we investigate the relation between entanglement and Bell nonlocality by using lots of randomly generated two-qubit states, and give out a constraint inequality relation between the two quantum resources. For studying the upper or lower boundary of the inequality relation, we discover maximally (minimally) nonlocal entangled states, which maximize (minimize) the value of the Bell nonlocality for a given value of the entanglement. Futhermore, we consider a special kind of mixed state transformed by performing an arbitrary unitary operation on werner state. It is found that the special mixed state's entanglement and Bell nonlocality are related to ones of a pure state transformed by the unitary operation performed on the Bell state. 
\end{abstract}
\maketitle

\section{Introduction}
Entanglement and Bell nonlocality are two core concepts that are used to describe non-classical correlations in quantum information. As an important quantum resource, entanglement is at the basis of many protocols in quantum
information \cite{w01}. For instance, remote state preparation \cite{w02,w03}, quantum teleportation \cite{w04}, super-dense coding \cite{w05}, quantum cryptographic key distribution \cite{w06}, quantum computation \cite{w07} and so on. The best performance of such tasks requires maximally entangled states, one of the most important entanglement manipulations is the entanglement purification or distillation \cite{w08,w09,w10,w11}. 

It is well known that Bell nonlocality is a sufficient form of quantum inseparabilities \cite{w29}. This nonlocal property is manifested explicitly by violating the different Bell-type inequalities, and it makes an essential role in better understanding of the subtle aspects of quantum mechanics \cite{w17,w19}. Some local quantum measurements reveal Bell nonlocality of bipartite states. However, their statistics of measurement outcomes can not be explained by a local hidden variable (LHV) model \cite{w16,w18}. This non-classical nature of quantum mechanics can be applied to device-independent quantum information processing \cite{w18}.

It is impossible that an entangled state can be expressed as a convex combination of separable states. And it is a fact that if a system has Bell nonlocality, then it must be entangled \cite{w20}. Historically, the violation of Bell inequality is usually used as a criterion for whether two qubits are entangled, for the inseparability of a two-qubit pure state corresponds to the violation of the Clauser-Horne-Shimony-Holt (CHSH) inequality, and vice versa \cite{w17}. However, for a two-qubit mixed state which are in practice the ones always encountered, this is not the case. Werner \cite{w21} proved initially that a general bipartite mixed state with entanglement does not violate any Bell-type inequalities. The violation of Bell inequality \cite{w16} has been recently confirmed by experiments which is not afflicted by detection and locality loopholes \cite{wa,wb,wc,wd}. And this violation of Bell inequality constitutes one of the most impressive confirmations of the nonlocal character of quantum theory.

Now that both entanglement and Bell nonlocality are used to describe quantum inseparabilities. As two vital quantum resources, we are more concerned about which domain the values of Bell nonlocality are limited in the values of entanglement. And what states the  boundary of Bell nonlocality located in entanglement region represents. In addition, we want to know evolution characteristics of the two quantum resources  by performing an arbitrary unitary operation on werner state. One difficulty in acquiring a general conclusion about the relation between entanglement and Bell nonlocality is to find a general incontrovertible methods that can compare entanglement and Bell nonlocality. Thus, the main goal of our research is how to obtain a universal relation between entanglement and Bell nonlocality.

The remainder of this article is organized as follows. In Sec. II, we review the quantification of entanglement and Bell nonlocality. In Sec. III, we investigate the relation between concurrence and Bell nonlocality by using lots of randomly generated two-qubit states. Interestingly, we discover an inequality relation between concurrence and Bell nonlocality. Based on the inequality, we discover maximally nonlocal entangled states and minimally nonlocal entangled states. In Sec. IV, we consider the relation between concurrence and Bell nonlocality of a special kind of mixed state, which is the state transformed by an arbitrary unitary operation from the werner state. In final, we end up our article with a brief conclusion.

\section{Preliminaries} \label{sec2}
Concurrence is usually used as a measure for entanglement of two-qubit states \cite{w01,w25}. For a two-qubit pure state $\left| \psi  \right\rangle $, its concurrence is defined as \cite{w26}
\begin{align}
C\left( {\left| \psi  \right\rangle } \right) = \left| {\left\langle {\psi }
	\mathrel{\left | {\vphantom {\psi  {\tilde \psi }}}
		\right. \kern-\nulldelimiterspace}
	{{\tilde \psi }} \right\rangle } \right|,
\end{align}
where $\left| {\tilde \psi } \right\rangle {\rm{ = (}}{\sigma _y} \otimes {\sigma _y})\left| {{\psi ^*}} \right\rangle $. Here $\left| {{\psi ^*}} \right\rangle $ is the complex conjugate of the pure state $\left| \psi  \right\rangle $ and ${\sigma _y}$ is the Pauli-y matrice. For a general two-qubit state $\rho $, its concurrence is defined by the convex-roof \cite{w11,w28} as follows
\begin{align}
C\left( \rho  \right) = \mathop {\min }\limits_{\left\{ {{q_n},\left| {{\varphi _n}} \right\rangle } \right\}} \sum\limits_n {{q_n}C\left( {\left| {{\varphi _n}} \right\rangle } \right)}.
\end{align}
The minimization is taken over all possible decompositions $\rho $ into pure states.
An analytic solution of concurrence can be calculated \cite{w26}
\begin{align}
C\left( \rho  \right) = \max\left\{ {0,{\rm{ }}\sqrt {{\lambda _1}}  - \sqrt {{\lambda _2}}  - \sqrt {{\lambda _3}}  - \sqrt {{\lambda _4}} } \right\},
\end{align}
where ${\lambda _n}$ ($n \in \{ 1,2,3,4\} $) are the eigenvalues, in decreasing order, of the non-Hermitian matrix $\rho \tilde \rho $. Here, the matrix $\tilde \rho $ has the following form
\begin{align}
\tilde \rho  = {\rm{(}}{\sigma _y} \otimes {\sigma _y}){\rho ^*}{\rm{(}}{\sigma _y} \otimes {\sigma _y}),
\end{align}
where the matrix ${\rho ^*}$ is the complex conjugate of the state $\rho $. In addition, with respect to a two-qubit Bell diagonal state ${\rho _{BD}}$, its concurrence can be given by
\begin{align}
	C\left( {{\rho _{BD}}} \right) = \max \{ 0,{\rm{ }}2{\lambda _{\max }}\left( {{\rho _{BD}}} \right) - 1\},
\end{align}
where ${\lambda _{\max }}\left( {{\rho _{BD}}} \right)$ is the maximum eigenvalue of the state ${\rho _{BD}}$.

Besides, Bell inequality violation in quantum mechanics tells us that quantum correlations are quite different from classical correlations. In the case of two-qubit states, the CHSH inequality \cite{w29} is a well-known Bell inequality and has the important property that an arbitrary two-qubit pure state violates the CHSH inequality if only and if it is entangled. Considering the Hilbert space $H = {C^2} \otimes {C^2}$, the Bell-operator associated with the Bell inequality can be given by
\begin{align}
{B_{CHSH}} = \vec a \cdot \vec \sigma  \otimes \left( {\vec b + \vec b'} \right) \cdot \vec \sigma  + \vec a' \cdot \vec \sigma  \otimes \left( {\vec b - \vec b'} \right) \cdot \vec \sigma ,
\end{align}
where $\vec a$, $\vec a'$ and $\vec b$, $\vec b'$ are unit vectors describing the measurements on sides $A$ and $B$, respectively. Here, $\vec \sigma  = \left( {{\sigma _x},{\rm{ }}{\sigma _y},{\rm{ }}{\sigma _z}} \right)$ is a vector made up of Pauli matrices. Then the Bell inequality can be expressed as
\begin{align}
\left| {{{\left\langle {{B_{CHSH}}} \right\rangle }_\rho }} \right| =\left| {{\rm{Tr}}\left( {\rho {B_{CHSH}}} \right)} \right| \le 2.
\end{align}
In terms of the Horodecki's theorem \cite{w29}, the maximum expected value of the Bell-operator for a general two-qubit state $\rho$ has the following form
\begin{align}
{B_{\max }}\left( \rho  \right) = \mathop {\max }\limits_{\left\{ {a,\vec a',b,\vec b'} \right\}} \left| {{{\left\langle {{B_{CHSH}}} \right\rangle }_\rho }} \right| = 2\sqrt {M\left( \rho  \right)} ,
\end{align}
where $M\left( \rho  \right) = {u_1} + {u_2}$ and ${u_i}{\rm{ }}\left( {i \in \{ 1,{\rm{ }}2\} } \right)$ are two larger eigenvalues of the symmetric matrix ${U_\rho } = T_\rho ^T{T_\rho }$ constructed from a correlation matrix ${T_\rho }$ and its transpose matrix $T_\rho ^T$. The real matrix ${T_\rho }$ is formed by the coefficients $ {\rm Tr}\left( {\rho {\sigma _m} \otimes {\sigma _n}} \right)$ ($m,{\rm{ }}n \in \left\{ {x,{\rm{ }}y,{\rm{ }}z} \right\}$). The Bell inequality can be violated if only and if $M\left( \rho  \right) > 1$ \cite{w29}. In order to make sure whether the Bell inequality is violated, we usually use Bell nonlocality to quantify the maximal violation of the Bell inequality. Following Ref. \cite{w30,w31}, we consider the Bell nonlocality $N\left( \rho  \right)$ has the following form
\begin{align}
N\left( \rho  \right) = \sqrt {\max \{ 0,{\rm{ }}M\left( \rho  \right) - 1\} }.
\end{align}

\section{Inequality relation between concurrence and Bell nonlocality} \label{sec4}
For a general two-qubit pure state $\left| \psi  \right\rangle $, the quantity $M\left( \left| \psi  \right\rangle  \right)$ can be given by \cite{w30}
\begin{align}
{M}\left( {\left| \varphi  \right\rangle } \right) = 1 + {C^2}\left( {\left| \varphi  \right\rangle } \right).
\end{align}
Therefore, for the pure state $\left| \varphi  \right\rangle$, the relation between concurrence $C\left( {\left| \varphi  \right\rangle } \right)$ and Bell nonlocality $N\left( {\left| \varphi  \right\rangle } \right)$  is 
\begin{align}
N\left( {\left| \varphi  \right\rangle } \right)=C\left( {\left| \varphi  \right\rangle } \right).
\end{align}

Eq. (11) shows that Bell nonlocality is equivalent to concurrence in a pure state system. However, for a general two-qubit mixed state $\rho$, the relation between concurrence $C\left( \rho  \right)$ and Bell nonlocality $N\left( \rho  \right)$ is intricate. It is well known that Bell-nonlocal states form a strict subset of entangled states \cite{w32}. In other words, non-local states must be entangled states, but entangled states are not necessarily non-local states.  However, we are more concerned about the mutual constraint between concurrence and Bell nonlocality, i.e., when we only know the values of concurrence, where is value-range of Bell nonlocality? In order to obtain the mutual constraint between the concurrence $C\left( \rho  \right)$ and the Bell nonlocality $N\left( \rho  \right)$, we investigate lots of randomly generated two-qubit states. The result shows that there is a constraint relation between concurrence and Bell nonlocality, which can be expressed by an inequality. The inequality can be expressed as follows (see Fig. 1)
\begin{align}
\sqrt {\max\{ 0,{\rm{ }}2{C^2}\left( \rho  \right) - 1\} }  \le N\left( \rho  \right) \le C\left( \rho  \right).
\end{align}

Eq. (12) reveals that Bell-nonlocal states form a strict subset of entangled states, and the Bell nonlocality is constrained by the concurrence. For the pure state $\left| \varphi  \right\rangle$, its concurrence $C\left( {\left| \varphi  \right\rangle } \right)$ is equal to its Bell nonlocality $N\left( {\left| \varphi  \right\rangle } \right)$. The relation satisfied in the case of pure states is the upper boundary reflected by Eq. (12). Thus, the state $\left| \varphi  \right\rangle$ is a maximally nonlocal entangled state. A natural confusion arises of whether there exist some typical mixed states that satisfy the upper or lower boundary of Eq. (12). With this confusion, we find two kinds of mixed states that satisfy the upper boundary and lower boundary of Eq. (12), respectively. And the two kinds of mixed states can be obtained in the following way. We place particle $A$ of the Werner state in the phase damped (PD) channel and the amplitude damped (AD) channel to obtain its evolutionary state, respectively. For the sake of discussion, we will analyze the two cases in two subsections. Before we do that, we need to know the relation between concurrence and Bell nonlocality for the Werner state.

\begin{figure}
	\centering
	\includegraphics[width=8cm]{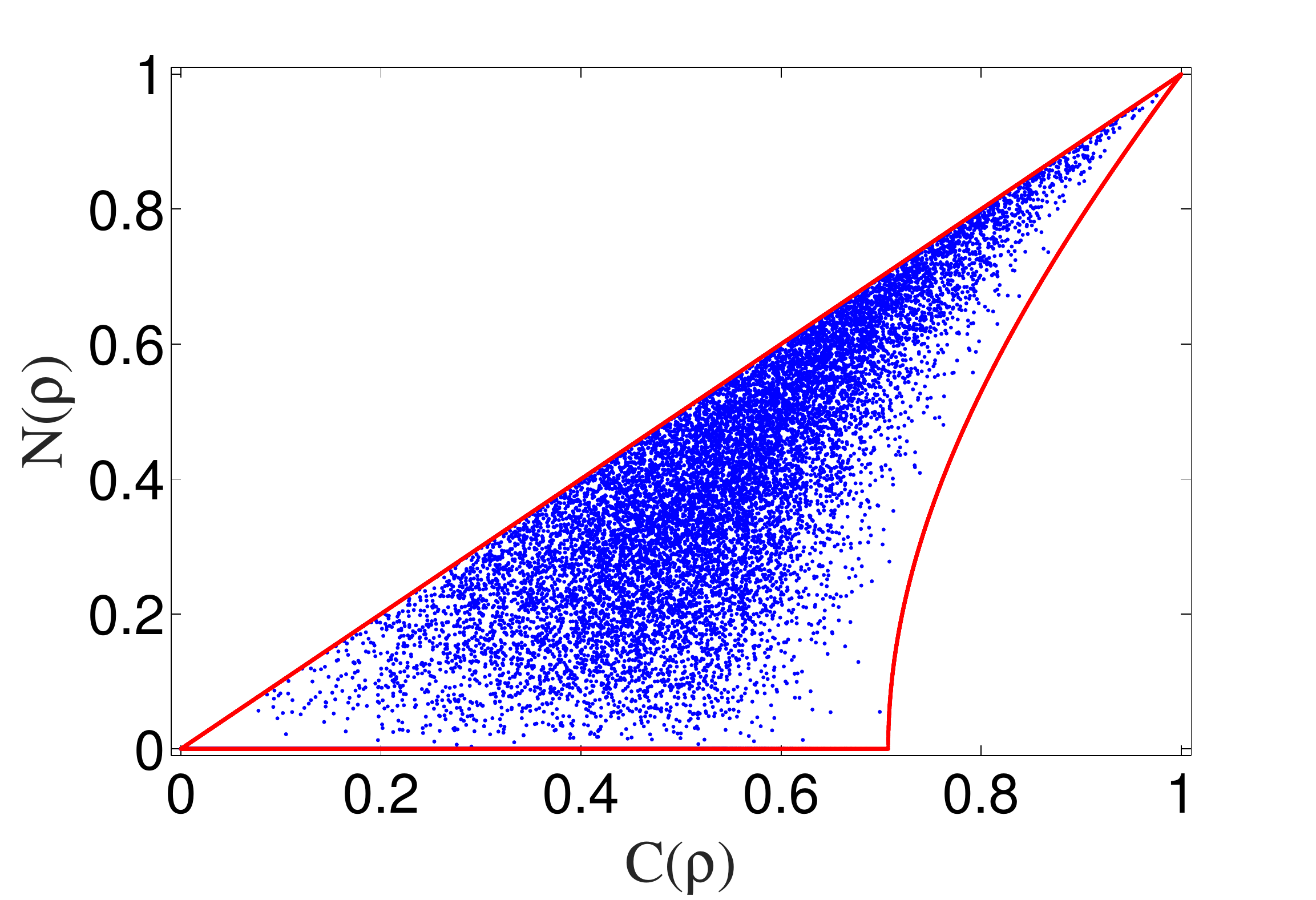}
	\caption{(Color online) Bell nonlocality  $N\left( \rho  \right)$ versus concurrence $C\left( \rho  \right)$ for two-qubit mixed states $\rho $. The upper and lower bounds (red line) are denoted by $N\left( \rho  \right) = C\left( \rho  \right)$ and $N\left( \rho  \right) = \sqrt {\max \{ 0,{\rm{ }}2{C^2}\left( \rho  \right) - 1\} } $,  respectively. The figure plots the Bell nonlocality $N\left( \rho  \right)$, along the Y axis, and the concurrence $C\left( \rho  \right)$, along the X axis, for $1.5 \times {10^5}$ randomly generated two-qubit states (there are $5 \times {10^4}$ randomly generated states of rank 2, 3, 4, respectively), by using a specific Matlab package.}
	\label{Fig1}
\end{figure}

We consider Werner state \cite{w21} as the initial state, which contains the maximal entangled pure state (Bell state) and the maximal mixed state, is defined as
\begin{align}
{\rho _W} = p\left| {{\varphi _B}} \right\rangle \left\langle {{\varphi _B}} \right| + (1 - p)\frac{\mathds{1} \otimes \mathds{1}}{4},
\end{align}
where the parameter $p$ is a real number in a closed interval $\left[ {0,1} \right]$. Here, the state $\left| {{\varphi _B}} \right\rangle  = \frac{1}{{\sqrt 2 }}\left( {\left| {00} \right\rangle  + \left| {11} \right\rangle } \right)$ is one of the Bell states and $\mathds{1}$ is the identity matrix whose order is $2$. The purity  of the state ${\rho _W}$ is $P\left( {{\rho _W}} \right) = \frac{{1 + 3{p^2}}}{4}$. When this state ${\rho _W}$ varies with the parameter $p$, we obtain the concurrence and Bell nonlocality with the following forms
\begin{align}
C\left( {{\rho _W}} \right) &= \max \{ 0,{\rm{ }}\frac{{3p - 1}}{2}\}, \nonumber \\ N\left( {{\rho _W}} \right) &= \sqrt {\max \{ 0,{\rm{ }}2{p^2} - 1\} }.
\end{align}
Hence, the relation between concurrence and Bell nonlocality for the Werner state ${\rho _W}$ can be given by
\begin{align}
N\left( {{\rho _W}} \right) = \frac{1}{3}\sqrt {\max \{ 0,{\rm{ }}8C\left( {{\rho _W}} \right) + 8{C^2}\left( {{\rho _W}} \right) - 7\} }.
\end{align}
Obviously, the Bell nonlocality is positively correlated with the concurrence for the Werner state ${\rho _W}$. When the concurrence $C\left( {{\rho _W}} \right)$  is greater than $\frac{{3\sqrt 2 -2 }}{{4}}$, the Bell inequality is violated.

\subsection{Evolutionary state corresponding to the PD channel}

We consider the evolutionary state ${\rho}$ which is formed by particle $A$ of the werner state ${\rho _W}$ going through the PD channel. And the state ${\rho}$ has the following concise form
\begin{align}
\rho  =& \sum\limits_{i = 0}^1 {{(K_i \otimes \mathds{1})}{\rho _W}(K_i^\dag \otimes \mathds{1})} \nonumber \\ =& p\sum\limits_{i = 0}^1 {{(K_i \otimes \mathds{1})}\left| {{\varphi _B}} \right\rangle \left\langle {{\varphi _B}} \right|(K_i^\dag \otimes \mathds{1}) }  \nonumber \\& + \frac{{1 - p}}{4}\sum\limits_{i = 0}^1 {{K_i}K_i^\dag \otimes \mathds{1} } \nonumber \\ =& p{\rho _{MNMS}} + (1 - p)\frac{\mathds{1} \otimes \mathds{1}}{4},
\end{align}
where ${K_0} = \left| 0 \right\rangle \left\langle 0 \right| + \varepsilon \left| 1 \right\rangle \left\langle 1 \right|$ and ${K_1} = \sqrt {1 - {\varepsilon ^2}} \left| 1 \right\rangle \left\langle 1 \right|$ ($\varepsilon  \in \left[ {0,1} \right]$) are the Kraus operators of the PD channel. Here, the state ${\rho _{MNMS}} = \sum\limits_{i = 0}^1 {{(K_i \otimes \mathds{1})}\left| {{\varphi _B}} \right\rangle \left\langle {{\varphi _B}} \right|(K_i^\dag \otimes \mathds{1}) }  $ is a maximally nonlocal mixed state. This is because the state ${\rho _{MNMS}}$ maximizes the value of the Bell nonlocality for a given value of the purity \cite{w33,w34}. For the evolutionary state $\rho $, it belongs to the Bell diagonal states ${\rho _{BD}}$. And the correlation matrix ${T_\rho }$ can be written as
\begin{align}
{T_\rho } = \left( {\begin{array}{*{20}{c}}
	{p\varepsilon }&0&0\\
	0&{ - p\varepsilon }&0\\
	0&0&p
	\end{array}} \right).
\end{align}
Hence, the Bell nonlocality $N\left( \rho  \right)$ of the state $\rho$ can be expressed as
\begin{align}
N\left( \rho  \right) = \sqrt {\max \{ 0,{\rm{ }}{p^2}\left( {1 + {\varepsilon ^2}} \right) - 1\} }.
\end{align}
For calculating the concurrence of this state $\rho $, we adopt Eq. (5) for concurrence of Bell diagonal states ${\rho _{BD}}$ to obtain the concurrence result of the state $\rho $, i.e.,
\begin{align}
C\left( \rho  \right) = \max \{ 0,{\rm{ }}p\varepsilon  - \frac{{1 - p}}{2}\}.
\end{align}
Hence, we acquire the mutual constraint (see Fig. 2) between concurrence and Bell nonlocality for the state ${\rho}$. It is worth noting that the upper and lower boundaries of the Fig. 2 correspond to the maximally nonlocal mixed state ${\rho _{MNMS}}$ and the Werner state ${\rho _W}$, respectively.

When the parameter $p$ is equal to 1, the state $\rho $ is reduced to the maximally nonlocal mixed state ${\rho _{MNMS}}$. Thus, its concurrence $C\left( {{\rho _{MNMS}}} \right)$ and Bell nonlocality $N\left( {{\rho _{MNMS}}} \right)$  can be given by
\begin{align}
C\left( {{\rho _{MNMS}}} \right) &= \varepsilon, \nonumber \\ N\left( {{\rho _{MNMS}}} \right) &= \varepsilon.
\end{align}
Eq. (20) shows that Bell nonlocality is equivalent to concurrence for the maximally nonlocal mixed state ${\rho _{MNMS}}$. Combining with Eq. (12), we obtain that the state ${\rho _{MNMS}}$ maximizes the value of the Bell nonlocality for a given value of the concurrence. Therefore, the state ${\rho _{MNMS}}$ is also a maximally nonlocal entangled state.

\begin{figure}
	\centering
	\includegraphics[width=7cm,height=5.2cm]{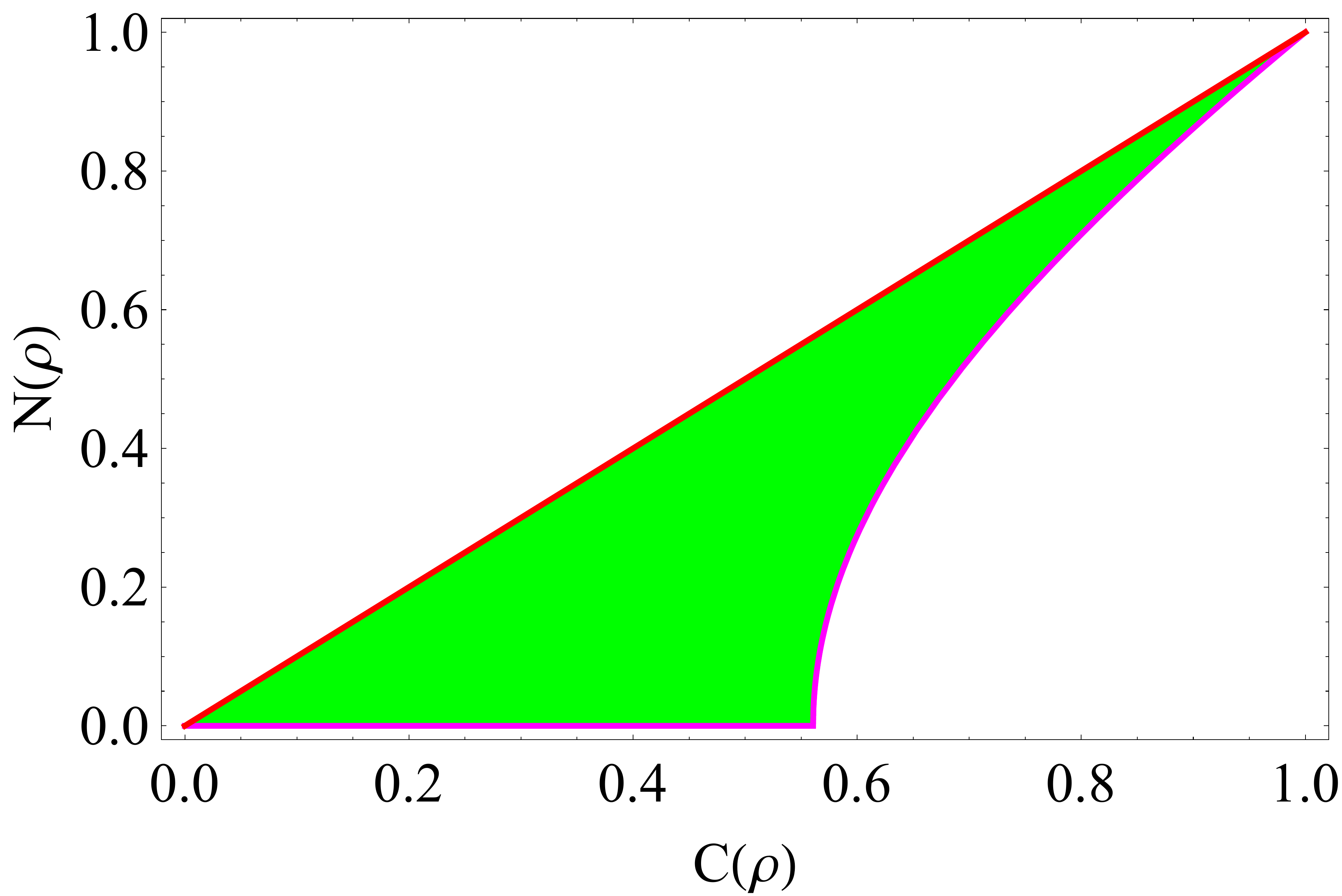}
	\caption{(Color online) Bell nonlocality  $N\left( \rho  \right)$ versus concurrence $C\left( \rho  \right)$ for the state $\rho$ corresponding to the PD channel.}
	\label{Fig2}
\end{figure}

\subsection{Evolutionary state corresponding to the AD channel}

We consider the evolutionary state ${\rho}$ which is formed by particle $A$ of the werner state ${\rho _W}$ going through the AD channel. And the state ${\rho}$ has the following concise form
\begin{align}
\rho= &\sum\limits_{i = 0}^1 {{(K_i \otimes \mathds{1})}{\rho _W}(K_i^\dag \otimes \mathds{1})} \nonumber \\ = & p\sum\limits_{i = 0}^1 {{(K_i \otimes \mathds{1})}\left| {{\varphi _B}} \right\rangle \left\langle {{\varphi _B}} \right|(K_i^\dag \otimes \mathds{1}) }  \nonumber \\&+ \frac{{1 - p}}{4}\sum\limits_{i = 0}^1 {{K_i}K_i^\dag \otimes \mathds{1} } \nonumber \\ =&p{\rho _{MNES}} + (1 - p){\rho _{NCMS}},
\end{align}
where ${\rho _{MNES}} = \sum\limits_{i = 0}^1 {{(K_i \otimes \mathds{1})}\left| {{\varphi _B}} \right\rangle \left\langle {{\varphi _B}} \right|(K_i^\dag \otimes \mathds{1}) }$ and ${\rho _{NCMS}} = \frac{1}{4}\sum\limits_{i = 0}^1 {{K_i}K_i^\dag \otimes \mathds{1}} $. Here, ${K_0} = \left| 0 \right\rangle \left\langle 0 \right| + \varepsilon \left| 1 \right\rangle \left\langle 1 \right|$ and ${K_1} = \sqrt {1 - {\varepsilon ^2}} \left| 0 \right\rangle \left\langle 1 \right|$ ($\varepsilon  \in \left[ {0,1} \right]$) are the Kraus operators of the AD channel. For the evolutionary state $\rho $, its correlation matrix ${T_\rho }$ can be written as
\begin{align}
{T_\rho } = \left( {\begin{array}{*{20}{c}}
	{p\varepsilon }&0&0\\
	0&{ - p\varepsilon }&0\\
	0&0&p\varepsilon ^2
	\end{array}} \right).
\end{align}
Hence, the Bell nonlocality $N\left( \rho  \right)$ of the state $\rho $ can be expressed as
\begin{align}
N\left( \rho  \right) = \sqrt{\max \{ 0,{\rm{ }}2{p^2}{\varepsilon ^2} - 1\}}.
\end{align}
For the calculation of the concurrence of this state $\rho $, we adopt Eq. (3) to compute the concurrence of the state $\rho $, i.e.,
\begin{align}
C\left( \rho  \right) = \max \{ 0,{\rm{ }}p\varepsilon  - \frac{{\sqrt {(1 - p)(2 - {\varepsilon ^2} - p{\varepsilon ^2})} }}{2}\varepsilon \}.
\end{align}

Hence, we get the mutual constraint (see Fig. 3) between concurrence and Bell nonlocality for the state ${\rho}$. It is worth noting that lower boundary of the Fig. 3 correspond to the state ${\rho _{MNES}}$, and upper boundary of the Fig. 3 correspond to the state ${\rho}$ at the case of $p= \varepsilon$.

When the parameter $p$ is equal to 0, the state $\rho $ is reduced to the  state ${\rho _{NCMS}}$. Here, the state ${\rho _{NCMS}}$ is a non-correlated mixed state. This is because its correlation function ${t_{ij}}\left( {{\rho _{NCMS}}} \right) =  {\rm Tr}\left( {{\rho _{NCMS}}{\sigma _i} \otimes {\sigma _j}} \right) = 0$. So, the state ${\rho _{NCMS}}$ is a non-entangled state. In particular, when the parameter $\varepsilon$ is equal to $1$, the state ${\rho _{NCMS}}$ is the maximum mixed state.

When the parameter $p$ is equal to 1, the state $\rho $ is reduced to the state ${\rho _{MNES}}$. Thus, its concurrence $C\left( {{\rho _{MNES}}} \right)$ and Bell's nonlocality $N\left( {{\rho _{MNES}}} \right)$  can be given by
\begin{align}
C\left( {{\rho _{MNES}}} \right) &= \varepsilon, \nonumber \\ N\left( {{\rho _{MNES}}} \right) &= \sqrt{\max \{ 0,{\rm{ }}2{\varepsilon ^2} - 1\}}.
\end{align}
Eq. (25) shows that when the concurrence $C\left( {{\rho _{MNES}}} \right)$ is greater than $\frac{1}{{\sqrt 2 }}$, the Bell inequality is violated for the state ${\rho _{MNES}}$. Combining with Eq. (12), we obtain that the state ${\rho _{MNES}}$ minimizes the value of the Bell nonlocality for a given value of the concurrence. Therefore, the state ${\rho _{MNES}}$ is a minimally nonlocal entangled state.

\begin{figure}
	\centering
	\includegraphics[width=7cm,height=5.2cm]{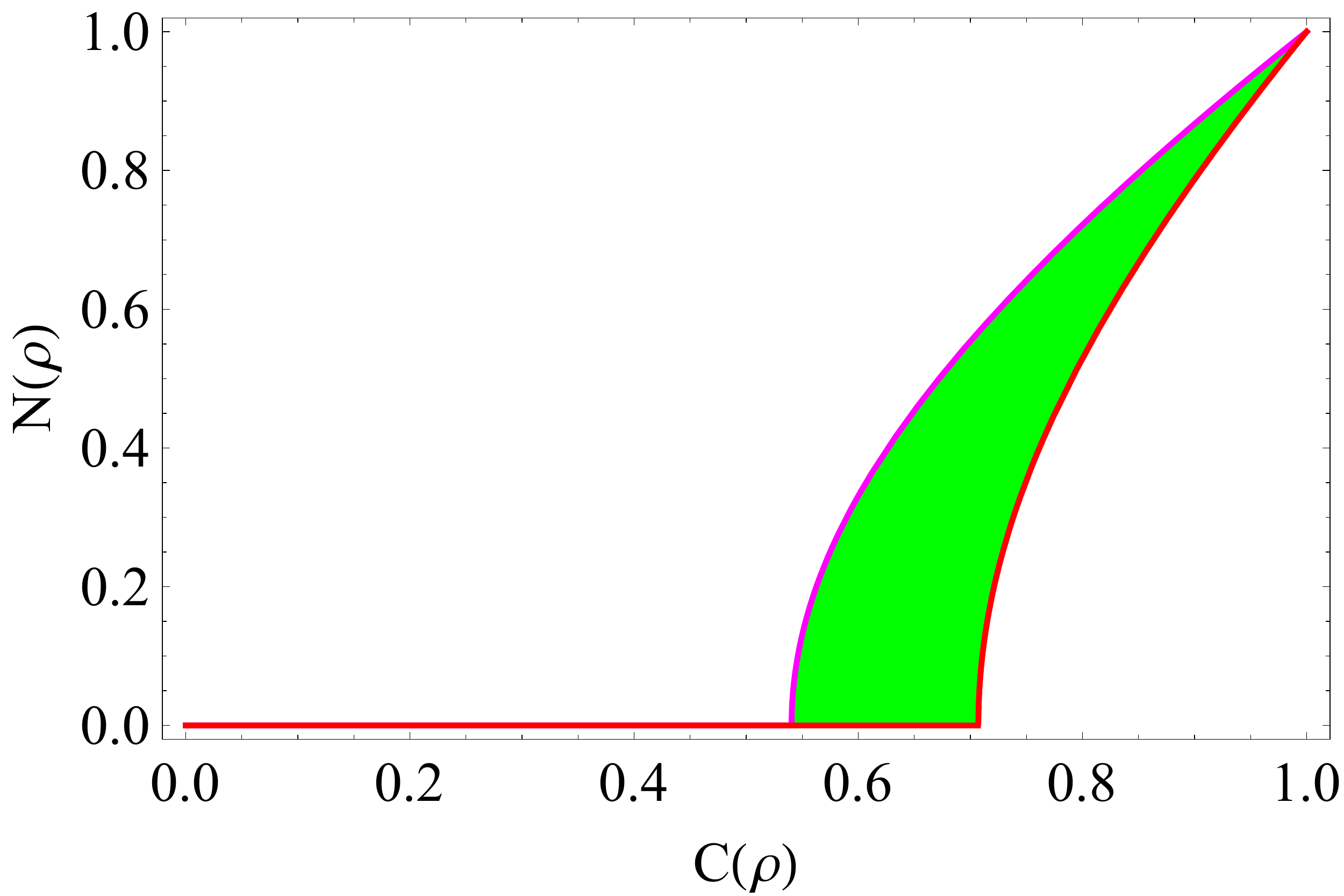}
	\caption{(Color online) Bell nonlocality  $N\left( \rho  \right)$ versus concurrence $C\left( \rho  \right)$ for the state $\rho$ corresponding to the AD channel.}
	\label{Fig3}
\end{figure}

\section{Concurrence and Bell nonlocality of a special kind of mixed state}

At the front, we have proposed the inequality relation between the concurrence and the Bell nonlocality for a general two-qubit state. For investigating the upper and  lower boundaries  of their relation between the concurrence and the Bell nonlocality, we send one qubit of werner state to go through PD and AD channel, respectively, and obtain the maxmally nonlocal entangled state and the minimally nonlocal entangled state. Next, we will study the  concurrence and the Bell nonlocality for the mixed state $\rho_{WU}$ which is the state transformed by performing an arbitrary unitary operation $U$ on the werner state $\rho_W$. In order to make the description simple, we consider that $U\left| {{\varphi _B}} \right\rangle $ is a pure state $\left| \varphi  \right\rangle $. And the mixed state $\rho_{WU}$ has the following form
\begin{align}
\rho_{WU} = U{\rho _W}{U^\dag } = p\left| \varphi  \right\rangle \left\langle \varphi  \right| + (1 - p)\frac{\mathds{1} \otimes \mathds{1}}{4}.
\end{align}
For the state $\rho_{WU}$, we obtain two properties about the concurrence $C\left( {{\rho _{WU}}} \right)$ and the Bell nonlocality $N\left( {{\rho _{WU}}} \right)$.

\textit{Property 1.} The Bell nonlocality $N\left( {{\rho _{WU}}} \right)$ of the state $\rho_{WU}$ is related to the Bell nonlocality ${N}\left( {\left| \varphi  \right\rangle } \right)$ of the pure state $\left| \varphi  \right\rangle $. And the correlation can be expressed as
\begin{align}
N\left( \rho _{WU}  \right) = \sqrt {\max \{ 0,{\rm{ }}{p^2}\left( {1 + {N^2}\left( {\left| \varphi  \right\rangle } \right)} \right) - 1\} }.
\end{align}

\textit{Proof of property 1.} The correlation function ${t_{ij}}\left( \rho_{WU}  \right)$ corresponding to the state $\rho_{WU}$ can be reduced as
\begin{align}
{t_{ij}}\left( \rho_{WU}  \right) &= {\rm Tr}\left( {\rho_{WU} {\sigma _i} \otimes {\sigma _j}} \right) \nonumber \\&= p {\rm Tr}\left( {\left| \varphi  \right\rangle \left\langle \varphi  \right|{\sigma _i} \otimes {\sigma _j}} \right) + \frac{{1 - p}}{4} {\rm Tr}\left( {{\sigma _i} \otimes {\sigma _j}} \right) \nonumber \\&= p{t_{ij}}\left( {\left| \varphi  \right\rangle } \right) + \frac{{1 - p}}{4} {\rm Tr}\left( {{\sigma _i}} \right) {\rm Tr}\left( {{\sigma _j}} \right) \nonumber \\&= p{t_{ij}}\left( {\left| \varphi  \right\rangle } \right).
\end{align}
Thus, the value ${M}\left( \rho_{WU}  \right)$ of the state $\rho_{WU} $ is closely related to the value ${M}\left( {\left| \varphi  \right\rangle } \right)$ of the pure state $\left| \varphi  \right\rangle $, i.e.,
\begin{align}
{M}\left( \rho_{WU} \right) = p^2{M}\left( {\left| \varphi  \right\rangle } \right).
\end{align}
Combining with Eqs. (9), (10) and (11), we obtain Eq. (27).

\textit{Property 2.} The concurrence $C\left( {{\rho _{WU}}} \right)$ of the state $\rho_{WU}$ is related to the concurrence ${C}\left( {\left| \varphi  \right\rangle } \right)$ of the pure state $\left| \varphi  \right\rangle $. And the correlation can be expressed as
\begin{align}
C\left( \rho_{WU}  \right) = \max \{ 0,{\rm{ }}pC\left( {\left| \varphi  \right\rangle } \right) - \frac{{1 - p}}{2}\}.
\end{align}

\textit{Proof of property 2.} The non-Hermitian matrix ${\rho _{WU}}{\tilde \rho _{WU}}$ can be given by
\begin{align}
{\rho _{WU}}{\tilde \rho _{WU}} = H + \frac{{{{\left( {1 - p} \right)}^2}}}{{16}} \mathds{1} \otimes \mathds{1},
\end{align}
where $H = {p^2}\left\langle {\varphi } \mathrel{\left | {\vphantom {\varphi  {\tilde \varphi }}}
	\right. \kern-\nulldelimiterspace} {{\tilde \varphi }} \right\rangle \left| \varphi  \right\rangle \left\langle {\tilde \varphi } \right| + \frac{{p\left( {1 - p} \right)}}{4}\left( {\left| \varphi  \right\rangle \left\langle \varphi  \right| + \left| {\tilde \varphi } \right\rangle \left\langle {\tilde \varphi } \right|} \right)$.
Here, we do not directly calculate the eigenvalues of non-Hermitian matrix ${\rho _{WU}}{\tilde \rho _{WU}}$, since its eigenvalue-equation are too complicated. According to some properties of matrix-rank, the rank $R\left( H \right)$ of the matrix $H$ is related to the ranks $R\left( {\left| \varphi  \right\rangle } \right)$ and $R\left( {\left| {\tilde \varphi } \right\rangle } \right)$. And the relation can be given by 
\begin{align}
R\left( H \right) \le& R\left( {\frac{p}{2}\left\langle {\varphi }
	\mathrel{\left | {\vphantom {\varphi  {\tilde \varphi }}}
		\right. \kern-\nulldelimiterspace}
	{{\tilde \varphi }} \right\rangle \left| \varphi  \right\rangle \left\langle {\tilde \varphi } \right| + \frac{{1 - p}}{4}\left| \varphi  \right\rangle \left\langle \varphi  \right|} \right) \nonumber \\ & + R\left( {\frac{p}{2}\left\langle {\varphi }
	\mathrel{\left | {\vphantom {\varphi  {\tilde \varphi }}}
		\right. \kern-\nulldelimiterspace}
	{{\tilde \varphi }} \right\rangle \left| \varphi  \right\rangle \left\langle {\tilde \varphi } \right| + \frac{{1 - p}}{4}\left| {\tilde \varphi } \right\rangle \left\langle {\tilde \varphi } \right|} \right) \nonumber \\ \le & R\left( {\left| \varphi  \right\rangle } \right) + R\left( {\left| {\tilde \varphi } \right\rangle } \right)=2.
\end{align}
It showes that at least two of the eigenvalues of the matrix $H$ are 0. Therefore, two eigenvalues ($\lambda _3$ and $\lambda _4$) of the non-Hermitian matrix ${\rho _{WU}}{\tilde \rho _{WU}}$ can be given by
\begin{align}
{\lambda _3} = {\lambda _4} = \frac{{{{\left( {1 - p} \right)}^2}}}{{16}}.
\end{align}
According to the equations $\sum\nolimits_{n = 1}^4 {{\lambda _n}}  = Tr\left( {\rho _{WU} \tilde \rho _{WU} } \right)$ and $\prod _{n = 1}^4{\lambda _n} = Det\left( {\rho _{WU} \tilde \rho _{WU} } \right)$, we obtain the other two eigenvalues ($\lambda _1$ and $\lambda _2$) satisfy the following equations
\begin{align}
{\lambda _1} + {\lambda _2} &= {p^2}{C^2}\left( {\left| \varphi  \right\rangle } \right) + \frac{{\left( {1 + 3p} \right)\left( {1 - p} \right)}}{8}, \nonumber \\
{\lambda _1}{\lambda _2} &= {\left[ {\frac{{\left( {1 + 3p} \right)\left( {1 - p} \right)}}{{16}}} \right]^2}.
\end{align}
Combining with Eqs. (3), (33) and (34), we get that the concurrence $C\left( {{\rho _{WU}}} \right)$ of the state $\rho _{WU}$ can be expressed as
\begin{align}
	C\left( {{\rho _{WU}}} \right) &= \max \left\{ {0,\sqrt {{\lambda _1}}  - \sqrt {{\lambda _2}}  - \sqrt {{\lambda _3}}  - \sqrt {{\lambda _4}} } \right\} \nonumber \\ &= \max \{ 0,\sqrt {{\lambda _1} + {\lambda _2} - 2\sqrt {{\lambda _1}{\lambda _2}} }  - \frac{{1 - p}}{2}\}  \nonumber \\ &= \max \left\{ {0,pC\left( {\left| \varphi  \right\rangle } \right) - \frac{{1 - p}}{2}} \right\}.
\end{align}

According to the \textit{Property 1} and \textit{Property 2}, we obtain that the amount of entanglement of the state $\rho _{WU}$ can not exceed that of the original Werner state, and the amount of Bell nonlocality of the state $\rho _{WU}$ can not exceed that of the original Werner state. It shows that Bell inequality can not be violated at the case of $p\le\frac{{1}}{\sqrt{2}}$. Therefore, for the state $\rho _{WU}$, Bell inequality will be violated if and only if its concurrence $C\left( {{\rho _{WU}}} \right)$ is greater than ${C_p} = \sqrt {1 - {p^2}}  - \frac{{1 - p}}{2}$, where $p>\frac{{1}}{\sqrt{2}}$.

\section{Conclusion} \label{sec6}
In this paper, we investigate the relation between concurrence and Bell nonlocality by using lots of randomly generated two-qubit states. We propose an inequality relation between entanglement and Bell nonlocality. For obtaining the corresponding state of upper or lower boundary of the inequality relation more accurately, we send one qubit of werner state to go through PD or AD channel and obtain maximally (minimally) nonlocal entangled states, which maximize (minimize) the value of the Bell nonlocality for a given value of the entanglement. Here, the maximally nonlocal entangled states include two types, one of which is two-qubit pure states and the other of which is the maximally nonlocal mixed states. Futhermore, we study the evolution property of werner state about its entanglement and Bell nonlocality, where a special kind of mixed state can be produced by performing an arbitrary unitary operation on the werner state.  And we find that the special mixed state's entanglement and Bell nonlocality are related to ones of a pure state, which can be transformed by putting the unitary operation on the Bell state. Bell nonlocality have been at the center of an active and intense research activity in the theory and experiment of quantum information science. The results presented about entanglement and Bell nonlocality will provide one an appropriate choice for more effectively utilizing the two quantum resources in the quantum information tasks. 

\section*{Acknowledgements} 
This work was supported by the National Science Foundation of China under Grant Nos. 11575001 and 61601002, Anhui Provincial Natural Science Foundation (Grant No. 1508085QF139) and Natural Science Foundation of Education Department of Anhui Province (Grant No. KJ2016SD49).

\bibliographystyle{plain}

\begin{thebibliography}{99}
	\bibitem{w01} R. Horodecki, P. Horodecki, M. Horodecki, and K. Horodecki, Quantum entanglement, Rev. Mod. Phys. {\bf 81}, 865 (2009).
	\bibitem{w02} A. K. Pati, Minimum classical bit for remote preparation and measurement of a qubit, Phys. Rev. A {\bf 63}, 014302 (2000).
	\bibitem{w03} C. H. Bennett, D. P. Di Vincenzo, P. W. Shor, J. A. Smolin, B. M. Terhal, and W. K.Wootters, Remote state preparation, Phys. Rev. Lett. {\bf 87}, 077902 (2001).
	\bibitem{w04} C. H. Bennett, G. Brassard, C. Crpeau, R. Jozsa, A. Peres, and W. K. Wootters, Teleporting an unknown quantum state via dual classical and Einstein-Podolsky-Rosen channels, Phys. Rev. Lett. {\bf 70}, 1895 (1993).
	\bibitem{w05} C. H. Bennett and S. J. Wiesner, Communication via one- and two-particle operators on Einstein-Podolsky-Rosen states, Phys. Rev. Lett. {\bf 69}, 2881(1992).
    \bibitem{w06} A. Ekert, Quantum cryptography based on Bell's theorem, Phys. Rev. Lett. {\bf 67}, 661 (1991)
	\bibitem{w07} A. Ekert and R. Jozsa, Quantum computation and Shor's factoring algorithm, Rev. Mod. Phys. {\bf 68}, 733 (1996)
	\bibitem{w08} N. Gisin, Hidden quantum nonlocality revealed by local filters, Phys. Lett. A {\bf 210}, 151 (1996).
	\bibitem{w09} M. Horodecki, P. Horodecki, and R. Horodecki, Inseparable Two Spin-$\frac{1}{2}$ Density Matrices Can Be Distilled to a Singlet Form, Phys. Rev. Lett. {\bf 78}, 574 (1997).
	\bibitem{w10} C. H. Bennett, G. Brassard, S. Popescu, B. Schumacher, J. Smolin, and W. K. Wootters, Purification of Noisy Entanglement and Faithful Teleportation via Noisy Channels, Phys. Rev. Lett. {\bf 76}, 722 (1996).
	\bibitem{w11} C. H. Bennett, D. P. Di Vincenzo, J. Smolin, and W. K. Wootters, Mixed-state entanglement and quantum error correction, Phys. Rev. A {\bf 54}, 3824 (1996).
	\bibitem{w29}R. Horodecki, P. Horodecki, M. Horodecki, Violating Bell inequality by mixed spin-$\frac{1}{2}$ states: necessary and sufficient condition, Phys. Lett. A, {\bf 200}, 340 (1995).
	\bibitem{w17} J. F. Clauser, M. A. Horne, A. Shimony, and R. A. Holt, Proposed experiment to test local hidden-variable theories, Phys. Rev. Lett. {\bf 23}, 880 (1969).
	\bibitem{w19} M. A. Nielsen, I. L. Chuang, Quantum Computation and Quantum Information, Cambridge University Press, Cambridge (2000).
	\bibitem{w16} J. S. Bell, On the einstein podolsky rosen paradox, Physics {\bf 1}, 195 (1964).
	\bibitem{w18} N. Brunner, D. Cavalcanti, S. Pironio, V. Scarani, and S. Wehner, Bell nonlocality, Rev. Mod. Phys. {\bf 86}, 419 (2014).
	\bibitem{w20} N. Gisin, Bell's inequality holds for all non-product states, Phys. Lett. A {\bf 154}, 201 (1991).
	\bibitem{w21} R. F. Werner, Quantum states with Einstein-Podolsky-Rosen correlations admitting a hidden-variable model, Phys. Rev. A {\bf 40}, 4277 (1989).
	\bibitem{wa} B. Hensen et al., Loophole-free Bell inequality violation using electron spins separated by 1.3 kilometres, Nature {\bf 526}, 682 (2015).
	\bibitem{wb} L. K. Shalmet al., Strong loophole-free test of local realism, Phys. Rev. Lett.  {\bf 115}, 250402 (2015).
	\bibitem{wc} M. Giustina et al., Significant-loophole-free test of Bell's theorem with entangled photons, Phys. Rev. Lett. {\bf 115}, 250401 (2015).
	\bibitem{wd} W. Rosenfeld, D. Burchardt, R. Garthoff, K. Redeker, N. Ortegel, M. Rau, and H. Weinfurter, Event-ready Bell test using entangled atoms simultaneously closing detection and locality loopholes, Phys. Rev. Lett. {\bf 119}, 010402 (2017).
	\bibitem{w25} O. Ghne and G. Tth, Entanglement detection, Phys. Rep. {\bf 474}, 1 (2009).
	\bibitem{w26} W. K. Wootters, Entanglement of Formation of an Arbitrary State of Two Qubits, Phys. Rev. Lett. {\bf 80}, 2245 (1998).
	\bibitem{w28} H. Barnum and N. Linden, Monotones and Invariants for Multi-particle Quantum States, J. Phys. A {\bf 34}, 6787 (2001).
	\bibitem{w30} A. Miranowicz, Violation of Bell inequality and entanglement of decaying Werner states, Phys. Lett. A {\bf 327}  272 (2004).
	\bibitem{w31} W. C. Ma, S. Xu, J. D. Shi and L. Ye, Quantum correlation versus Bell-inequality violation under the amplitude damping channel, Phys. Lett. A {\bf 379} 2802-2807 (2015).
	\bibitem{w32} M. T. Quintino, T. Vertesi, D. Cavalcanti, R. Augusiak, M. Demianowicz, A. Ac\'{i}n, and N. Brunner, Inequivalence of entanglement, steering, and Bell nonlocality for general measurements, Phys. Rev. A {\bf 92}, 032107 (2015).
	\bibitem{w33} J. Batle and M. Casas, Nonlocality and entanglement in qubit systems, J. Phys. A {\bf 44}, 445304 (2011).
	\bibitem{w34} J. Svozil\'{i}k, A. Vall\'{e}s, J. Pe\v{r}ina, Jr., and J. P. Torres, Revealing Hidden Coherence in Partially Coherent Light, Phys. Rev. Lett. {\bf 115}, 220501 (2015).	
\end{thebibliography}

\end{document}